\documentclass[epj]{svjour}
\usepackage{graphics}

\begin{document}
\title{Bloch oscillations of a soliton in a molecular chain}
\author{V.D.Lakhno
\thanks{\emph{E-mail: lak@impb.psn.ru}} \and A.N.Korshunova
}

\institute{Institute of Mathematical Problems of Biology, Russian
Academy of Sciences, Pushchino, Moscow Region, Russian Federation}

\date{Received: 13 November 2006}

\abstract{ The paper presents results of numerical experiments
simulating Bloch oscillations of solitons in a deformable
molecular chain in a constant electric field. By the example of a
homogeneous polynucleotide chain it is shown that the system under
consideration can demonstrate complicated dynamical regimes when
at the field intensities less than a certain critical value, a
soliton as a whole exhibits oscillations, while at the field
intensities exceeding the threshold, a soliton turns to a breather
which oscillates. It is shown that the motion of a charge in a
deformable chain is infinite as contrasted to that in a rigid
chain.
} 
\maketitle It is well known that an electron occurring in an ideal
rigid periodic molecular chain or in a solid state superlattice
exhibits Bloch oscillations in response to a constant electric
field \cite{1}-\cite{5}. In an external time-periodic field,
motion of a charge along a rigid chain can be both infinite and
finite (dynamical localization) \cite{6}-\cite{10}. In a
deformable crystal chain the role of an external field is played
by oscillations of the lattice nodes which can be presented as
superposition of plane travelling waves, or phonons. In this case
the motion of an electron along the chain is thought to be
infinite since the electron scatters on phonons and Bloch
oscillations do not take place \cite{11}.

It is common knowledge that in quasi-one-dimensional molecular
chains interaction of an electron with lattice oscillations is not
weak. Therefore we cannot safely assume that the electron wave
function goes off phase (in view of  scattering of the electron on
phonons)  and  Bloch oscillations fail.

To clear up this point we consider the case when a charge placed
in a molecular chain transits to a soliton state as a result of
interaction with lattice oscillations. This occurs, for example,
in homogeneous polynucleotide chains where the charge motion is
described by\, Holstein Hamiltonian\, in\, which \,each site
\,presents  a nucleotide pair \,considered \, as \,a harmonical
oscillator \,\cite{12}-\cite{14}:
\begin{eqnarray}\label{1}
  \nonumber\hat{H} &=& \hat{H}_h+\hat{T}_k+\hat{U}_p\,, \\
  \hat{H}_h &=& \nu\sum_{n=1}^{N}(a_n^+a_{n-1}+a_n^+a_{n+1})+\!\!\sum_{n=1}^{N}\alpha_na_n^+a_n\,, \\
  \nonumber\hat{T}_k &=& \!\!\sum_{n=1}^{N}\frac{\hat{P_n^2}}{2M},\;\;
  \hat{U}_p=\!\!\sum_{n=1}^{N}k\frac{q_n^2}{2},\;\;\alpha_n\!=\alpha'q_n+n\hbar\omega_B.
\end{eqnarray}
Here $\hat{H}_h$  - is a Hamiltonian of a charged particle,
$a_n^+,a_n$ - are operators of creation and annihilation of the
charge on site $n$, $\nu$  - is the matrix element of the
transition from the $n$ - th site to the $n\pm1$ - site,
$\alpha_n$ - is the energy of the particle at the $n$ - th site,
$\hbar\omega_B=e\mathcal{E}\!a$, where $\mathcal{E}$~- is the
intensity of the electric field, $e$ - is the electron charge, $a$
- is the distance between neighboring bases. $\hat{T}_k$ - is an
operator of the kinetic energy of sites, $\hat{U}_p$ - is the
potential energy of sites, $\hat{P}_n$ - is an impulse operator
canonically conjugated to the displacement $q_n$, $M$ - is the
effective mass of the site, $k$~- is an elastic constant,
$\alpha'$ - is the particle-site displacement coupling constant.

We can pass on to semi classical description of the wave function
of the system $|\Psi(t)\rangle$  as an expansion over coherent
states:
\begin{equation}\label{2}
\!\!\!\!|\Psi(t)\rangle\!=\!\!\!\sum_{n=1}^{N}\!b_n(t)a_n^+\!\!\exp\Bigl\{\!-\frac{i}{\hbar}\!\sum_{j}\!\!\left[\beta_j(t)\hat{P}_j-\pi_j(t)q_j\right]\Bigr\}|0\rangle,
\end{equation}
where $|0\rangle$  - is the vacuum wave function and the
quantities $\beta_j(t)$ and $\pi_j(t)$ satisfy the relations:
\begin{equation}\label{3}
\langle\Psi(t)|q_n|\Psi(t)\rangle=\beta_n(t),\quad
\langle\Psi(t)|\hat{P}_n|\Psi(t)\rangle=\pi_n(t).
\end{equation}
Dynamical equations for the quantities $b_n(t)$ and $\beta_n(t)$
resulting from (\ref{1}) - (\ref{3})  have the form : 
\begin{equation}\label{4}
i\hbar\dot{b}_n=\alpha_nb_n+\nu(b_{n-1}+b_{n+1}),
\end{equation}
\begin{equation}\label{5}
M\ddot{\beta}_n=-\gamma\dot{\beta}_n-k\beta_n-\alpha'|b_n|^2.
\end{equation}

\begin{figure}
 \begin{center}
\resizebox{0.49\textwidth}{0.9\textheight}{%
  \includegraphics{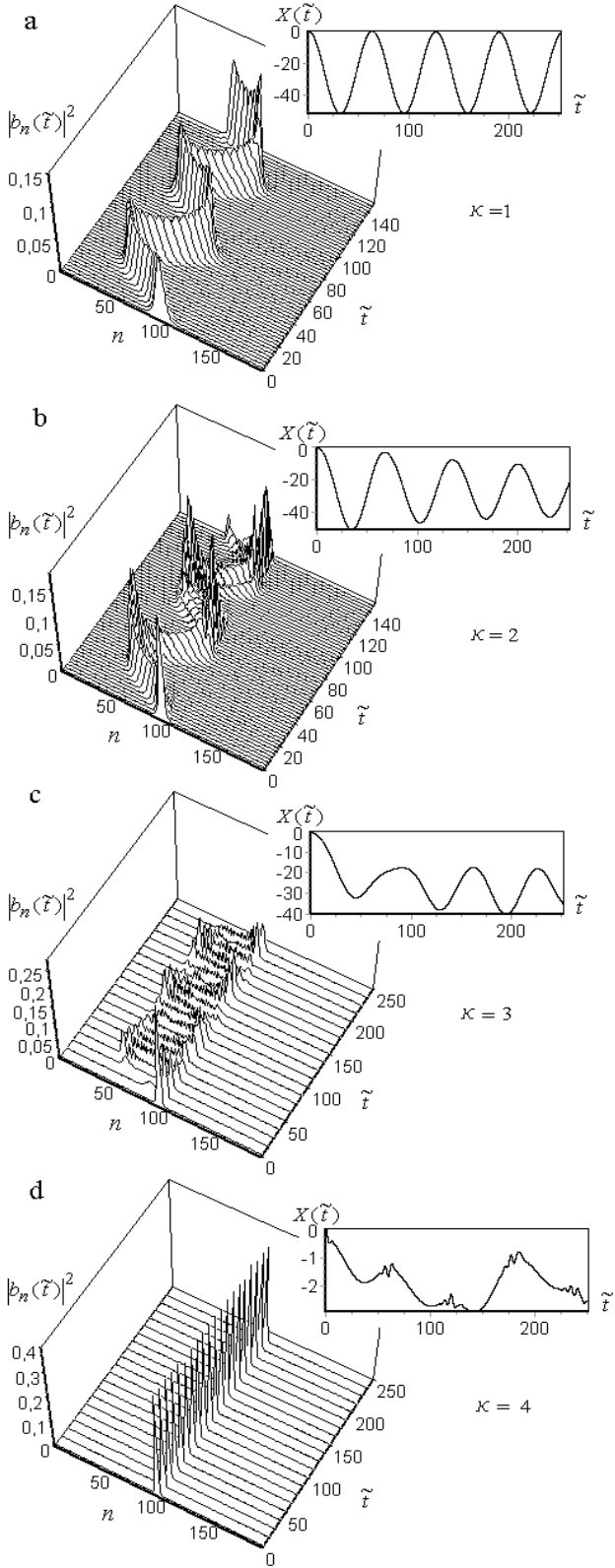}
}
 \end{center}
\caption{Oscillatory motions of a soliton for some values of
parameter $\kappa (\kappa=1,2,3,4)$ at the electric field intensity
$E~\!\!=~\!\!0.1$. The length of the homogeneous nucleotide chain is $N=201$,
$\widetilde{t}=t/\tau$, $\tau=10^{-14}sec$, $\widetilde{\omega}\,'=0.006$, $\widetilde{\omega}=0.01$,
$\eta=1.276$.
$(\widetilde{\omega}\,'=\omega\,'\tau$, $\omega\,'=\gamma/M=6\cdot10^{11}sec^{-1}$ \cite{12}, $\widetilde{\omega}=\omega\tau)$.}
\label{fig:1}
\end{figure}

Equations (\ref{4}) are Schr\"{o}dinger equations where $b_n$ is
the amplitude of the particle localization at the $n$ - th site.
Equations (\ref{5}) are classical motion equations describing
dynamics of nucleotide pairs with regard for dissipation, where
$\gamma$ is friction coefficient. We believe that a semiclassical
description in which motion of a charge along a chain is described
by quantum motion equations (\ref{4}) and motion of individual
nucleotides is presented by classical motion equations (\ref{5})
is valid in view of a large nucleotide mass ($\approx$~300 proton
mass).

In the case of a rigid chain, when $\alpha'=0$ the solution of the
system (\ref{4}), (\ref{5}) will be \cite{15}, \cite{16}:
\begin{eqnarray}\label{6}
  \nonumber b_n(t) &=& \!\!\!\sum_{m=-\infty}^\infty \!\!b_m(0)(-i)^{n-m}e^{-i(n+m)\omega_Bt/2}J_{n-m}\left(\xi(t)\right), \\
  \xi(t) &=&
  \frac{4\nu}{\hbar\omega_B}\sin\!\!\left(\frac{\omega_Bt}{2}\right),
\end{eqnarray}
$J_n(x)$ - is Bessel function of the first kind. Solution
(\ref{6}) corresponds to Bloch oscillations of a particle in the
chain affected by an electric field for which the particle's
centre mass:
\begin{equation}\label{7}
X(t)=\sum_{n=1}^N|b_n(t)|^2na ,
\end{equation}
demonstrates periodic oscillations at the frequency of $\omega_B$:
\begin{eqnarray}\label{8}
  \nonumber X(t) &=& X(0)+\frac{2a\nu}{\hbar\omega_B}\left|S_0\right|\bigl(\cos\theta_0-\cos(\omega_Bt+\theta_0)\bigr), \\
  S_0 &=&\!\!\!\sum_{m=-\infty}^\infty \!\!b_m^*(0)b_{m-1}(0)=\left|S_0\right|e^{i\theta_0}, \\
  \nonumber X(0) &=& a\!\!\!\sum_{m=-\infty}^\infty \!\!\!m|b_m(0)|^2,
\end{eqnarray}
where $a$ is the distance between neighboring nucleotides, which
for DNA is equal to 3.4$\mathring{A}$.

For $\alpha'\neq0$   in the absence of an electric field, a
stationary solution of equations (\ref{4}), (\ref{5}) corresponds
to a localized state of a soliton type. To study the evolution of
a soliton state in an electric field we will use an initial charge
density distribution such that:
\begin{eqnarray}\label{9}
|b_n(0)|=\frac{\sqrt{2}}{4}\cosh^{-1}\Bigl(\frac{\kappa(n-n_0)}{4\eta}\Bigr),\\
\nonumber n_0=\frac{\,N}{2}+1,\quad \eta=\frac{\nu\tau}\hbar,\quad
\kappa=\frac{\tau\alpha'^2}{k\hbar}.
\end{eqnarray}
Initial values of $x^0$ and $y^0$ $\left(b_n=x_n+iy_n\right)$ for
$\nu>0$ have the form:
\begin{equation}\label{10}
\!\!x_n^0=|b_n(0)|(-1)^n\!\!\left/\!\sqrt{2}\right. , \,
\,\,y_n^0=|b_n(0)|(-1)^{n+1}\!\!\left/\!\sqrt{2}\right. ,
\end{equation}
which corresponds to the ground state of a particle in the absence
of an electric field \cite{13},\cite{14}.

\begin{figure}
\begin{center}
\resizebox{0.4\textwidth}{0.3\textheight}{%
  \includegraphics{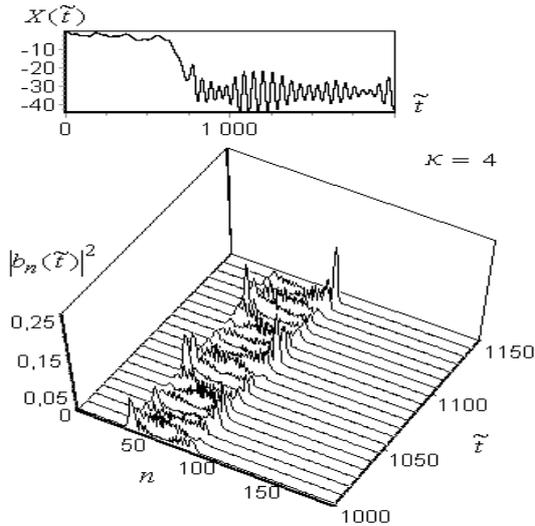}}
\end{center}
\caption{Transition of a soliton into a breather for $\kappa=4$.
\mbox{\qquad\qquad}(Fig.1 d) at large times.)} \label{fig:2}
\end{figure}

Fig.1 shows the results of the solution of equations (\ref{4}),
(\ref{5}) for some values of parameter $\kappa$, responsible for
the intensity of the charge interaction with lattice oscillations
at the electric field intensity $E=\mathcal{E}ea\tau/\hbar=0.1$,
$\widetilde{\omega}~=~\omega\tau=0.01$, \,$\eta=1.276$. Here the
values of parameters $\omega$ and $\eta$ are the same as in work
\cite{12}, and $\tau=10^{-14}$sec. In dimensional units these
parameter values correspond to $\mathcal{E}~=~1.94\cdot10^5V/cm$,
$\omega=\sqrt{k/M}=10^{12}sec^{-1}$, $\nu=0.084eV$. The parameter
of electron-phonon strength $\kappa=4$, which in dimensional units
corresponds to $\alpha'=0.13eV/\mathring{A}$ is the same as in
\cite{12}. This value is close to that used by other authors (in
\cite{18} $\alpha'$ was found to be
$\alpha'\approx0.23eV/\mathring{A}$).

It is seen from Fig.\ref{fig:1} a) that in the presence of an
electric field, a soliton executes periodical motions, coming back
to the point where the soliton center mass initially occurred.
This oscillatory motion corresponds to Bloch oscillations with a
period of $T=2\pi/\omega_B$. The total amplitude of the
oscillations $L$ is close to that determined from the solution of
linear problem (\ref{6}) and is written as: $\Delta Wa/E$, where
$\Delta W=4\eta$ stands for the width of the conductivity band
equal to $\Delta W\tau/\hbar$ in dimension form. For the parameter
values presented above $L\approx51a$ (with the soliton
characteristic size $\approx10a$).

Fig.\ref{fig:1} b)-d) shows evolution of the dynamical behavior of
a soliton at the initial stages of the motion as parameter
$\kappa$ increases. After a lapse of time Bloch oscillations
restore (restored Bloch oscillations are not given in
Fig.\ref{fig:1} b)-d)).

In the case of strong electric fields presented in
Fig.\ref{fig:1}, a soliton executing Bloch oscillations with time
turns to a breather oscillating at Bloch frequency
(Fig.\ref{fig:2}). At rather large values of $\kappa$, a breather
can arise from the initial soliton state immediately, i.e.
by-passing the phase of Bloch oscillations as a whole.

\begin{figure}
\begin{center}
\resizebox{0.37\textwidth}{0.22\textheight}{%
  \includegraphics{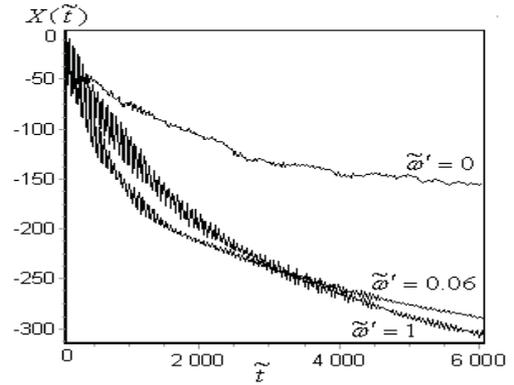}}
\end{center}
\caption{Function $X(\widetilde{t})$ for various values of
$\widetilde{\omega}\,'$.} \label{fig:3}
\end{figure}

Without going into details of nonstationary regimes of the
particle motion in the cases under consideration we will restrict
ourselves to mere qualitative description of the picture. It has
been observed that the case of a deformable chain $(\alpha'\neq0)$
differs qualitatively from the limiting case of a rigid chain
$(\alpha'=0)$ in that at finite $\alpha'$ the quantity $X(t)$
given by (\ref{7}) grows infinitely at $t\rightarrow\infty$
(Fig.\ref{fig:3}). This result could have been guessed from the
already mentioned analogy between the influence of a periodic
external electric field on a particle and oscillations of phonons.
Quite nontrivial is the finding that under this influence, in the
case of a strong particle-phonons interaction, i.e. when a soliton
is formed, Bloch oscillations of the particle persist in the
electric field as oscillations of a soliton as a whole or a
breather, depending on the system parameters.

In conclusion it may be said that this picture of the charge
motion in a deformable molecular chain in a constant electric
field at zero temperature $T=0$ seems to be rather general: a
positive charge introduced in the chain will move along the field
executing Bloch oscillations. At finite temperatures a soliton or
breather state will break thus leading to failure of Bloch
oscillations. In this case motion of the charge over the chain
will be infinite along the lines of the field and have an ordinary
band character.

\end{document}